\documentclass[aps,pra,twocolumn,nofootinbib,floatfix,10pt]{revtex4-2}

\usepackage{amsmath,mathtools}
\usepackage{amssymb}
\usepackage{wasysym}
\usepackage{graphicx}
\usepackage{color,soul}
\usepackage{physics}
\usepackage{siunitx}
\usepackage{placeins}
\usepackage{dsfont}
\usepackage{float}
\usepackage[english]{babel}
\usepackage{blindtext}
\usepackage[english,nomargin,inline,marginclue,draft]{fixme}
\pdfpageheight\paperheight
\pdfpagewidth\paperwidth

\usepackage[colorlinks,linkcolor=blue,anchorcolor=blue,citecolor=blue,urlcolor=blue]{hyperref}
\usepackage{printlen}
\usepackage{ulem}
            
\fxusetheme{colorsig}
\FXRegisterAuthor{cg}{acg}{CG}  
\FXRegisterAuthor{th}{ath}{\color{blue}TH}  
\FXRegisterAuthor{ib}{aib}{\color{red}IB} 
\FXRegisterAuthor{sh}{ash}{\color{cyan}SH} 
\FXRegisterAuthor{db}{adb}{\color{green}DB} 
\FXRegisterAuthor{ps}{aps}{PS}
\makeatletter
\renewcommand*\FXLayoutInline[3]{%
  {\@fxuseface{inline}\ignorespaces{\color{fx#1}[#3: #2]}}}
\makeatother

\long\def\symbolfootnote[#1]#2{\begingroup%
\def\thefootnote{\fnsymbol{footnote}}\footnotetext[#1]{#2}\endgroup}

\def\nobreakbefore{%
  \relax\ifvmode\else
    \ifhmode
      \ifdim\lastskip > 0pt\relax
        \unskip\nobreakspace
      \else 
        \nobreakspace
      \fi
    \fi
  \fi
}
\let\oldcite\cite
\renewcommand\cite{\nobreakbefore\oldcite}

\begin{document}
\title{Self-Supervised Learning with Noisy Dataset for Rydberg Microwave Sensors Denoising}

\author{Zongkai Liu$^{1,2,\textcolor{blue}{\star},\textcolor{blue}{\S}}$}

\author{Qiming Ren$^{1,2,\textcolor{blue}{\S}}$}

\author{Wenguang Yang$^{1,2}$}

\author{Yanjie Tong$^{1,2}$}

\author{Huizhen Wang$^{1,2}$}

\author{Yijie Zhang$^{1,2}$}

\author{Ruohao Zhi$^{1,2}$}

\author{Junyao Xie$^{1,2}$}

\author{Mingyong Jing$^{1,2}$}

\author{Hao Zhang$^{1,2}$}

\author{Liantuan Xiao$^{1,2}$}

\author{Suotang Jia$^{1,2}$}

\author{Ke Tang$^{3}$}

\author{Linjie Zhang$^{1,2,\textcolor{blue}{\ddagger}}$}

\affiliation{$^1$State Key Laboratory of Quantum Optics Technologies and Devices, Institute of Laser Spectroscopy, Shanxi University, Taiyuan 030006, China .}

\affiliation{$^2$Collaborative Innovation Center of Extreme Optics, Shanxi University, Taiyuan, Shanxi 030006, China.}

\affiliation{$^3$Guangdong Provincial Key Laboratory of Brain-inspired Intelligent Computation, Department of Computer Science and Engineering, Southern University of Science and Technology, Shenzhen 518055, China.  }

\symbolfootnote[1]{lzk1997@sxu.edu.cn}
\symbolfootnote[3]{zlj@sxu.edu.cn}
\
\symbolfootnote[4]{Z.L and Q.R contribute equally to this work.}
\date{\today}

\begin{abstract}
We report a self-supervised deep learning framework for Rydberg sensors that enables single-shot noise suppression matching the accuracy of multi-measurement averaging. The framework eliminates the need for clean reference signals (hardly required in quantum sensing) by training on two sets of noisy signals with identical statistical distributions. When evaluated on Rydberg sensing datasets, the framework outperforms wavelet transform and Kalman filtering, achieving a denoising effect equivalent to 10,000-set averaging while reducing computation time by three orders of magnitude. We further validate performance across diverse noise profiles and quantify the complexity-performance trade-off of U-Net and Transformer architectures, providing actionable guidance for optimizing deep learning-based denoising in Rydberg sensor systems.

 
\end{abstract}

\maketitle

\section{Introduction}
Rydberg atom microwave sensors\cite{sedlacek2012microwave,10.1117/12.2309386,jing2020atomic,9374680,yuan2023quantum,10238372,ZHANG20241515} are promising for applied fields such as wideband communications and radar, thanks to their ultra-high microwave electric field sensitivity \cite{PhysRevApplied.21.L031003,11123636}, broad frequency response (DC to THz)\cite{anderson2017continuous,adams2019rydberg,Meyer_2020}, and International System of Units (SI)-traceable calibration\cite{SI-Traceable,Holloway2017ElectricFM,9363580,schlossberger2024rydberg}. However, their utility in open-environment deployments is limited by noise sources that submerge the signal, such as environmental electromagnetic interference, interatomic interaction noise, and thermal noise \cite{jing2020atomic,ZHANG20241515}.

Traditional denoising approaches are ill-suited to this constraint. Multi-measurement averaging suppresses random noise but requires hundreds of repetitions, sacrificing temporal resolution and rendering it incompatible with time-varying signals\cite{1164833,dantus2004experimental}. Wavelet transform (based on thresholding and signal matching)\cite{806084,PASTI199921,GrossmannMorlet+2006+126+139,HUIMIN20121354,8404418} and Kalman filtering (dependent on pre-defined state-space models)\cite{5311910} perform well for stationary noise but fail to stabilize under the non-stationary noise encountered in real-world Rydberg sensor operation.

Deep learning has emerged as an alternative for adaptive denoising\cite{yu2019deep}: recurrent neural networks (RNNs)\cite{grossberg2013recurrent,Caterini2018,WANG2022100296} , long short-term memory (LSTMs)\cite{LSTM,van2020review}and,1d-convolution neural network model\cite{7966039,8682194,ismail2019deep,electronics10010059,tonekaboni2021unsupervised,10.1145/3711896.3736899} leverage temporal dependencies to separate signals from noise (e.g., in speech processing \cite{6638947,sak2014long,weninger2015speech,9733937,wang2024towards}), while Transformers use self-attention to balance noise suppression and signal detail preservation\cite{vaswani2017attention,9710580,9937486,10443052}. 
However, a standard deep learning structure (e.g., autoencoders\cite{10.1145/1390156.1390294,lu2013,bank2023autoencoders}) requires paired noisy-clean signal datasets for supervised learning, where clean single-shot signals are experimentally inaccessible for Rydberg sensors.
Moreover,  theoretically constructed idealized clean data deviate from actual experimental conditions, which readily leads to model overfitting \cite{lever2016points,montesinos2022overfitting}.

\begin{figure*}
    \centering
    \includegraphics[width=1\linewidth]{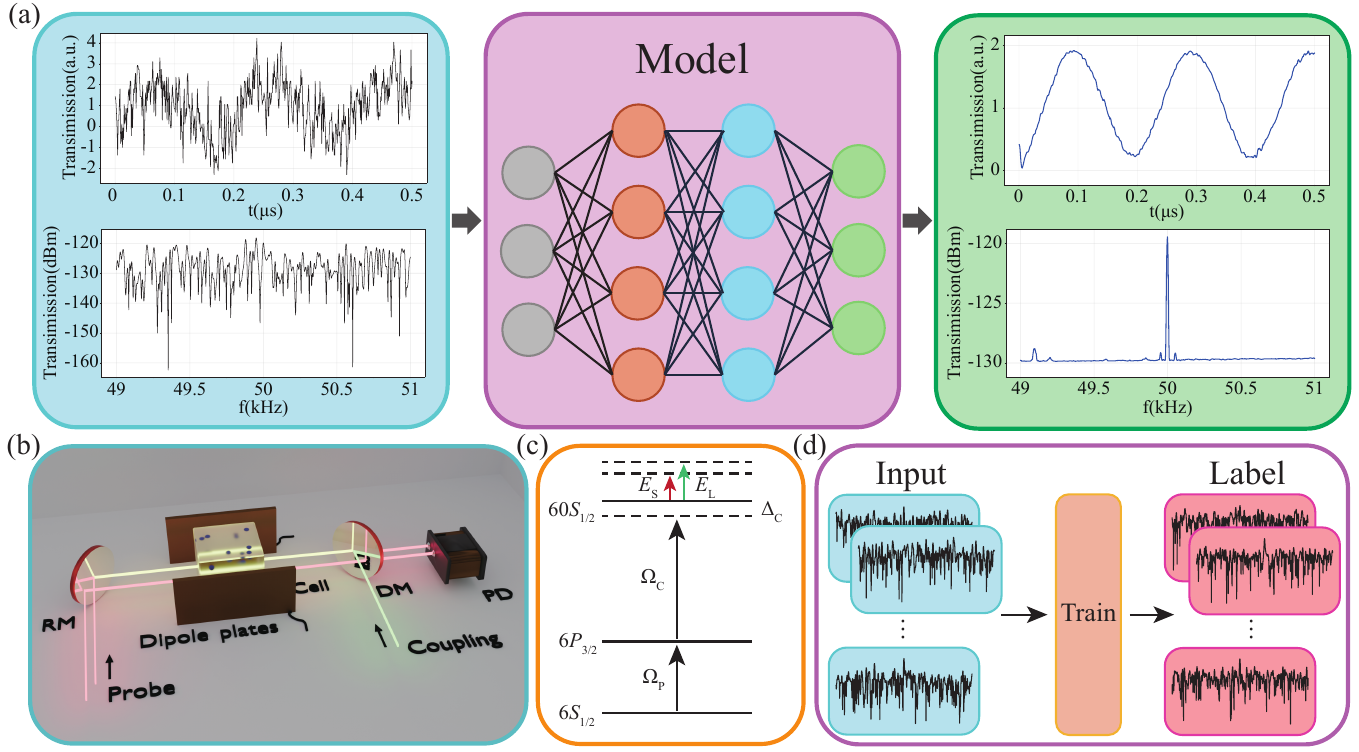}
    \caption{The process of model testing and training. (a) A trained model is capable of generating a clean signal from an input noisy signal including time (upper panel) and frequency domain (lower panel) data. The data are collected from experiment. 
    (b)Experimental setup. The probe light is split into two beams: one counter-propagates with the coupling light inside a cesium vapor cell, while the other outside the cell acts as a reference beam to enable balanced detection. The data set of time domain and frequency domain is collected by a oscilloscope and spectrum analyzer after the balanced detection. The signal electric field $E_{S}$ and local oscillator electric field $E_{L}$ interact with the atoms via parallel-plate electrodes. RM-reflect mirror, DM-dichroic mirror, PD-balanced photo-detector.
    (c) Energy level diagram. The probe light (Rabi frequency $\Omega_p$) and coupling light (Rabi frequency $\Omega_c$,  frequency detuning from resonant $\Delta_c$) excite atoms from the ground state $6S_{1/2}$ to the Rydberg state $60S_{1/2}$ via the intermediate state $6P_{3/2}$. The signal and local fields with amplitude $E_{S}$ and $E_{L}$ and frequency 63 MHz and 63.05 MHz act on the Rydberg state distribution. 
    (d) During training, the input data and their corresponding labels are independent measurement results of the microwave signal from the same Rydberg atom, differing only in identically independent distributed noise.}
    \label{fig:process}
\end{figure*}

Here, we develop a self-supervised deep learning framework to Rydberg atom microwave sensors denoising, shown in Fig.~\ref{fig:process} (a). It is validated on Rydberg sensing signals in both time and frequency domains and is compared with 10000-set averaging, Kalman filtering and wavelet filtering. It has three key advances for applied use: (1) it enables single-shot denoising with accuracy matching 10,000-set averaging, preserving temporal resolution for dynamic measurements; (2) it eliminates the need for clean reference signals by training on two sets of noisy signals with identical statistical distributions, resolving the labeled data scarcity in quantum sensing; (3) it reduces computation time by three orders of magnitude compared to multi-measurement averaging.  Finally, we quantify the complexity-performance trade-off of U-Net and Transformer architectures to guide model selection for specific sensor deployment needs. This work advances the practical utility of Rydberg sensors by bridging deep learning denoising with applied quantum sensing constraints.

\section{Methods}

\subsection{Experimental Setup}
A Rydberg atom microwave detection system was constructed to acquire noisy intermediate frequency (IF) signals for validating deep learning denoising methods. 
The experimental setup and energy level diagram are shown in Figs.~\ref{fig:process} (b) and (c), respectively.
Cesium atoms were confined in a room-temperature glass cell ($5\times5\times5$ $\mathrm{cm}^3$). 
A 852 nm probe beam (Rabi frequency \(\Omega_c/2\pi = 9.06\ \text{MHz}\)) excited atoms from \(|6S_{1/2}, F=4\rangle\) to \(|6P_{3/2}, F=5\rangle\), while a 510 nm coupling beam (Rabi frequency \(\Omega_c/2\pi = 0.83\ \text{MHz}\)) further excited atoms to the \(60S_{1/2}\) Rydberg state. 
The probe beam was split into two: one as a reference outside the cell, and the other counter-propagating with the coupling beam inside to form an electromagnetically induced transparency (EIT) configuration, where the coupling beam modulated the probe beam’s atomic transmission and absorption.
The probe beam including the reference are balanced detected by a differential detector (bandwidth 1 MHz and common mode rejection Ratio 40 dB) yielded spectral signals. 
Both beams’ frequencies were stabilized by an ultra-stable cavity (\(\text{finesse} = 2 \times 10^5\)) and locked to the aforementioned atomic transitions.

The experiment was conducted at room temperature in a microwave anechoic chamber lined with pyramidal foam absorbers, which provides an attenuation of  $<20$ dB for radio frequency (RF) fields at frequency $<100$ MHz incident normally on the material. 
Radio frequency (RF) electric field detection adopted an atomic heterodyne scheme: an arbitrary wave generator produced a 63 MHz signal under test (SUT) $E_S$ and a local oscillator (LO) signal $E_L$ frequency detuned from SUT by 50 kHz.
These signals were input to parallel brass electrodes (10 cm $\times$ 8 cm in size, 5 cm spacing) to generate a radio frequency (RF) field applied to the atomic system.
The atomic energy level is shifted by the SUT and local fields via the ac Stark shift, which is deduced from the probe transmission \cite{ZHANG20241515,liu2022highly}. 
The shift frequency $\delta$ is given by $ \delta=- \alpha |E_S+E_L|^2/2$,  where $\alpha$ is the scalar polarizability.
The beating between SUT $E_S$ and local fields $E_L$  generates  a 50 kHz IF signal whose amplitude correspond to the SUT’s.
The probe beam is detected by a balanced differential detector, which converts the light intensity into a voltage signal. 
This signal is then read out in the time domain using an oscilloscope and in the frequency domain using a spectrum analyzer for deep learning-based denoising. 
The spectrum analyzer is configured with a resolution bandwidth (RBW) of 10 Hz, 5 averaging cycles for spectrum acquisition, and a frequency sweep range of 49–51 kHz with a step size of 5 Hz.
In addition, different intensity of attenuation were applied to the IF signal to simulate open electromagnetic environments, where single-shot signals were submerged in noise but recoverable via multi-measurement averaging.

\subsection{Dataset Design and Training Strategy}

The primary goal of training the neural network $f_\theta(\cdot)$  is to learn an optimized weight parameter ($\theta^\star$) (encompassing the weights and biases) that minimizes the discrepancy between the network’s outputs and the experimental training labels (minimize the loss function $\mathcal{L}(\cdot)$). Mathematically, the training objective is formalized as a constrained minimization problem, and the optimization process is as follows: 

\begin{equation}
\theta^\star = \arg\min_{\theta} \frac{1}{N} \sum_{i=1}^{N} \mathcal{L}\left( f_\theta(x_{\text{train},i}), y_{\text{train},i} \right),
\end{equation}
with the size of training dataset $N $, $x_{\text{train}}=x+n_1$, $y_{\text{train}}=y+n_2$, where $x_{\text{train}}$ and $y_{\text{train}}$ denote two independent experimental measured data, which consist of ideal clean signals $x$, $y$ and noise $n_1$, $n_2$ drawn from the independent identical  noise distribution, i.e., $n_1 , n_2\in \text{Noise}$. 
The distinction between self-supervised learning and supervised learning lies in their training labels. 
For supervised learning (e.g., autoencoders\cite{9053925,lu2013}), the training label is an ideal, noise-free reference $y_{\text{train}}=y$ (such references are rarely obtainable in sensing scenarios). 
In contrast, the label used in self-supervised learning (in this work) corresponds to another measurement outcome same condition to training data $y_{\text{train}}=y+n_2$, which eliminates the need for clean reference. 
The noise $n_1 , n_2$  including environmental electromagnetic interference, interatomic interaction noise, and thermal noise, etc \cite{jing2020atomic}.
To adjust the effect of the noise on the signal, we applied an attenuation on the  IF signal from the scenario where the IF is above the noise floor to it is submerged by noise but is recognizable after 10000 averaging.
After training converges (defined as the training loss remaining stable for 10 consecutive epochs) the optimized weights $\theta^{\star}$ of the deep learning layers are obtained. 
The prediction process leverages unseen test inputs to output, formulated as
\begin{equation}
\hat{y}_{\text{test}} = f_{\theta^\star}(x_{\text{test}}) ,
\end{equation}
where  $x_{\text{test}}$ is independent of  $x_{\text{train}}$ to ensure unbiased generalization evaluation, the prediction result $\hat{y}_{\text{test}}$ is compared to the 10000 set average of the origin data.

Specifically, a Transformer-based end-to-end model was built, along with a dedicated dataset and training protocol, with key details below.
In uniform experiments, 10,000 sets of independent and identically distributed (i.i.d.) noisy spectral/time-domain data (1000 consecutive points each set) were collected and partitioned into training, labeling, and testing sets at a 4:4:2 ratio.
Specifically, the training set includes measurement result data \(x_{\text{train}}\) (dimension: 4000, 1000, 1), and independent another set of  measurement results \(y_{\text{train}}\) (dimension:  4000, 1000, 1),  while the test set consists of  \(x_{\text{test}}\) (dimension:  2000, 1000, 1).
Here $x_{(\cdot)}$ and $y_{(\cdot)}$ denote the training data and their corresponding labels, respectively; the first dimension of the data represents the number of sets, the second dimension denotes the number of consecutive points, and the third dimension is a requirement of the deep learning model.
The unlabeled testing data \(x_{\text{test}}\)  was tested against the average of 10,000 original data sets.

Next, to improve the stability of the training and avoid magnitude bias, each data point was standardized: \(x' = \frac{x - \mu}{\sigma},\)where $x$ = original data, \(\mu\) = mean, \(\sigma\) = standard deviation, and \(x'\) = standardized data. 
The data is then fed into a deep learning model for training, whose architecture is shown in Appendix \ref{sec:model}.

\begin{figure*}
    \centering
    \includegraphics[width=1\linewidth]{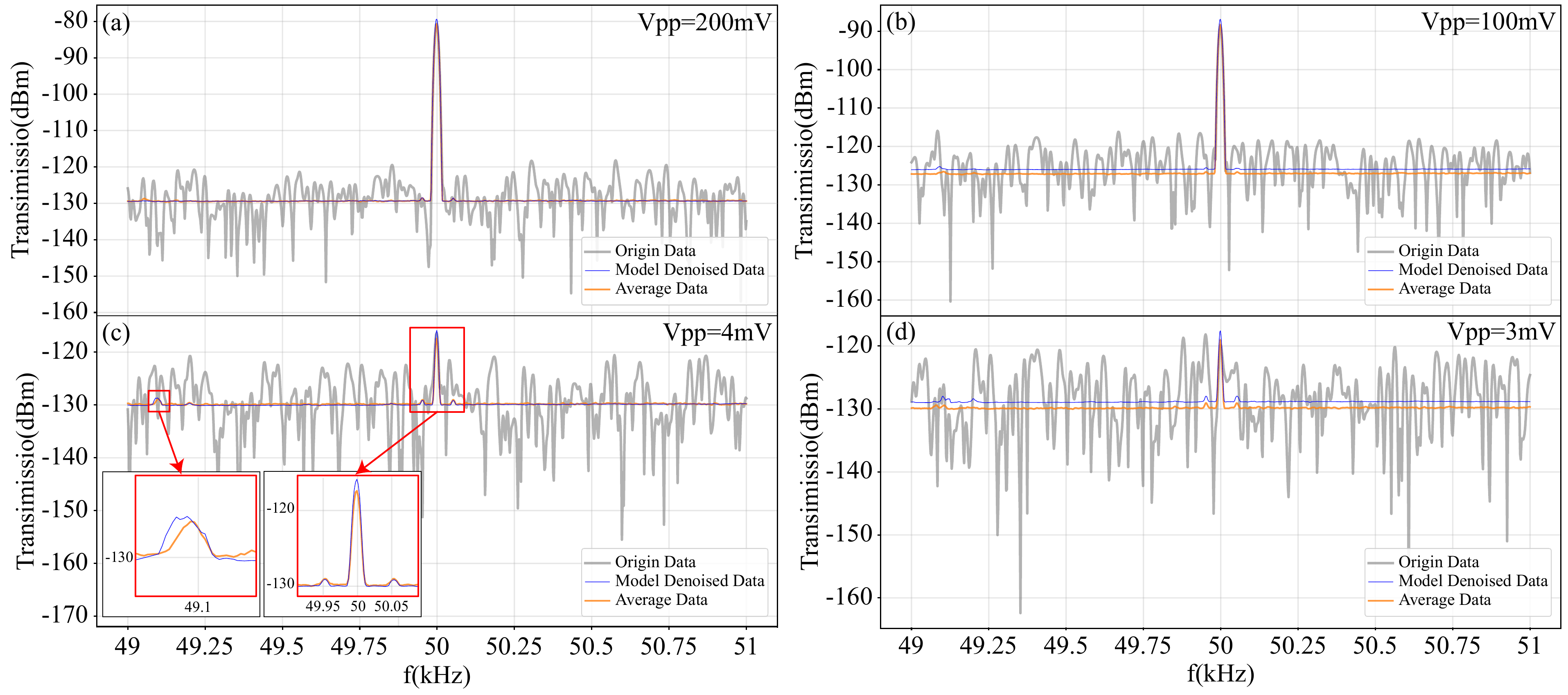}
    \caption{Intermediate frequency (IF) frequency-domain signals under different attenuation levels. IF Signal intensities are labeled in the top-right corner of Subfigures (a)–(d).  Gray curves: Single-shot measurements of probe beam transmission signals (attenuating from above noise floor to noise floor level). Green curves: Averaged results of 10,000 measurements (noise-free ground truth, not used in model training), showing noise elimination with only IF signal retained (plus IF sidebands and ~49 kHz weak signals under weak attenuation in (c)–(d)). Blue curves: Denoising results of Transformer based model (trained solely on noisy signals), achieving performance comparable to multi-measurement averaging and revealing IF signal, IF sidebands, and ~49.1 kHz weak signals. The two red boxes in (c) are zoomed-in views of the averaged and deep learning denoising results at frequencies of 50 kHz and 49.1 kHz, respectively, to illustrate the retained submerged signal.}
    \label{fig:frequency}
\end{figure*}

\subsection{Why the Training Process Reduces Noise in Noisy Datasets}

To clarify the noise reduction mechanism of the proposed training process, we analyse the optimization of the objective function combined with the statistical properties of noise and the law of large numbers (LLN), as detailed below (adopting MSE loss \(\mathcal{L}(a,b)=(a-b)^2\) for simplicity, extendable to other loss functions).

First, we assume that the noise \(n_1, n_2 \in \text{Noise}\) satisfies two general properties: (1)Zero mean: \(\mathrm{E}[n_1] = \mathrm{E}[n_2] = 0\) (no systematic bias); (2)Independence on clean signals: \(\mathrm{E}[x \cdot n_1] = \mathrm{E}[y \cdot n_2] = 0\) (noise does not carry signal-related information).
The training goal is to minimize the empirical loss:
\begin{equation}
    \min_\theta J(\theta) = \min_\theta \frac{1}{N}\sum_{i=1}^N \mathcal{L}\left(f_\theta(x_{\text{train},i}), y_{\text{train},i}\right).
\end{equation}
Substituting \(x_{\text{train},i} = x_i + n_{1,i}\) and \(y_{\text{train},i} = y_i + n_{2,i}\), where $x_{\text{train}}$ and $y_{\text{train}}$ are independent measurement results  corresponding to the microwave signal of the same Rydberg atom differ only in the noise, $x$ and $y$ are clean data.
The objective function becomes:
\begin{equation}
    J(\theta) = \frac{1}{N}\sum_{i=1}^N \left[ f_\theta(x_i + n_{1,i}) - (y_i + n_{2,i}) \right]^2.
\end{equation}

For large $N$, the empirical risk converges to the expected risk by LLN:
\begin{equation}
    J(\theta) \xrightarrow{N \to \infty} \mathrm{E}\left[ \left( f_\theta(x + n_1) - (y + n_2) \right)^2 \right].
\end{equation}
Expanding the expectation term and applying the noise assumptions:
\begin{equation}
    \mathrm{E}\left[ \cdots \right] = \mathrm{E}\left[ \left( f_\theta(x + n_1) - y \right)^2 \right] + \sigma_{n2}^2,
\end{equation}
where the cross term vanishes (\(\mathrm{E}[n_2] = 0\) and independence), and \(\mathrm{E}[n_2^2] = \sigma_{n2}^2\) (constant noise variance, irrelevant to \(\theta\)). 
Since \(\sigma_{n2}^2\) is fixed, minimizing \(J(\theta)\) is equivalent to minimizing the signal fitting error 
\begin{equation}
    \mathrm{E}\left[ \left( f_\theta(x + n_1) - y \right)^2 \right].
\end{equation}
To achieve this, the model must learn a noise-agnostic mapping, $
    f_\theta(x + n_1) \rightarrow y$.
    
Consequently, random fluctuations of \(n_1\) and \(n_2\) are averaged by LLN (for large $N$), and the model spontaneously suppresses noise by focusing on the clean $x-y$ relationship rather than fitting noise.
This explains why the training process reduces noise in noisy datasets.

\section{Results}

\subsection{Denoising Performance on Frequency-Domain Signals}

The deep learning model enables denoising quantum sensing data for both time-domain and frequency-domain signals. 
Fig.~\ref{fig:frequency} presents its denoising results on IF frequency-domain signals measured by the spectrum analyzer, comparing to the averaged result of 10,000 repeated datasets with i.i.d. noise. 
From Figs.~\ref{fig:frequency}  (a-d), the amplitude of SUT on the electrodes is gradually attenuated from 200 mV to 3 mV, which is used to simulate the scenario where the data is gradually submerged by noise. 
Since the IF signal arises from the beating of the SUT signal and the local signal, and the frequency of the local signal is adjustable to track the SUT, the frequency of the IF signal is fixed at 50 kHz.
Without loss of generality, this configuration can be generalized to SUTs with other frequencies.

\begin{figure}
    \centering
    \includegraphics[width=1\linewidth]{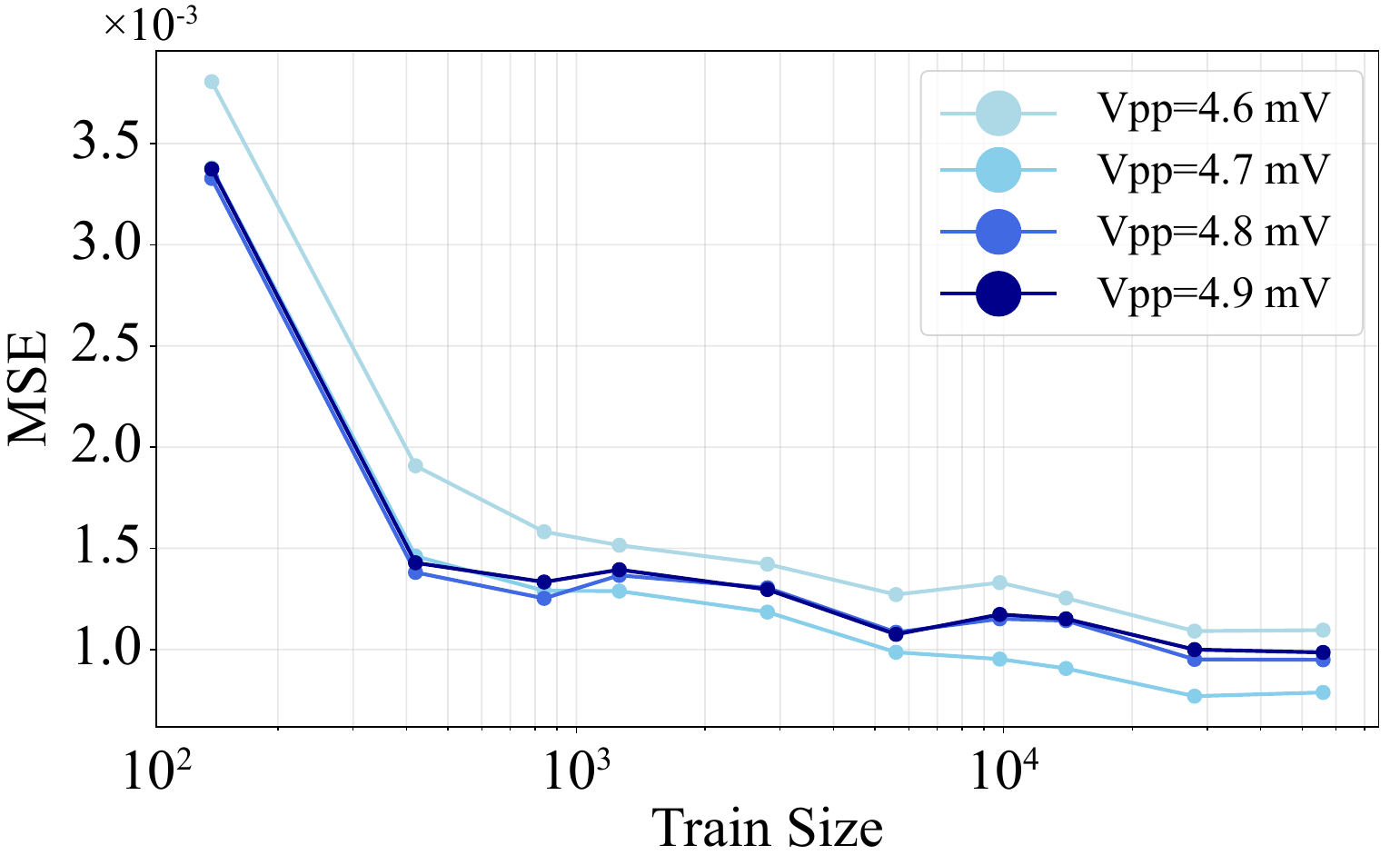}
    \caption{ MSE depends on the size of the training dataset in frequency domain. This panel presents the MSE between the denoised results of the deep learning model and the results of 10000-set averaging, under different voltages applied to the dipole plates from 4.6 mV to 4.9 mV. As the training dataset size increases, the MSE decreases.}
    \label{fig:size}
\end{figure}

For strong-amplitude signals (Figs.~\ref{fig:frequency} (a)–(b)), the model restores noisy data to a level comparable to the 10,000-measurement average, while fully preserving the 50 kHz intermediate frequency (IF) signal.  
The model outperformed multi-measurement averaging in temporal efficiency: measurement time was reduced by three orders of magnitude (1 hour for 10,000 measurements + averaging
vs. 0.7 second for deep learning denoising).
For spectral data with 40 dB attenuation in amplitude (Figs.  \ref{fig:frequency} (c)–(d)), the original signal is completely submerged in noise,however, the model still recovers the IF signal and retains its characteristic components. 
Notably, non-noise components are retained, as show in Fig.\ref{fig:frequency} (c) red box and insets: around the 50 kHz IF signal (two symmetric small peaks) and in the 49 to 49.25 kHz range (small peaks), signals emerge from noise in both the averaged and model denoised results. 
These are likely characteristic signals, indicating that the model can identify not only post-averaging signals far above the noise floor but also those close to it and submerged by it. 
The deep learning model successfully recognizes these characteristic signals that are consistent with those of the averaged results. 
This consistency not only validates the model’s ability to capture meaningful patterns, but also confirms that the denoising process is not a spurious artifact of the deep learning framework.
In other frequency regions, the denoised results are flatter than the average, showing superior noise suppression in noise-dominated ranges.

Denoising performance was quantified via mean squared error (MSE, against the 10000-measurement average): 
\begin{equation}
    \text{MSE} = \frac{1}{N}\sum_{i=1}^{N} \left(y_i - \bar{y}_i\right)^2,
\end{equation}
where \(N\) is the number of data points, \(y_i\) is the denoised output, and \(\bar{y}_i\) is the corresponding 10000-measurement average reference serving as the ground truth. 
In Fig. \ref{fig:size}, the number of training data is varied from 100 to $5\times10^4$ for peak-to-peak voltage Vpp ranging from 4.6 mV to 4.9 mV.
After each training process, the MSE is calculated  to evaluate how well the denoised results align with the averaging result.
It decrease significantly when the number of training data points is over 500,  indicating that a moderate increase in training data volume can drastically boost denoising efficacy, and reaches a  plateau as the number of training data points approaches $5\times10^4$.

In summary, the model achieves noise suppression performance comparable to that of multi-measurement averaging.
It uniquely enables the extraction of noise-submerged signals and the identification of potential characteristic signals from single-shot data.

\begin{figure}
    \centering
    \includegraphics[width=1\linewidth]{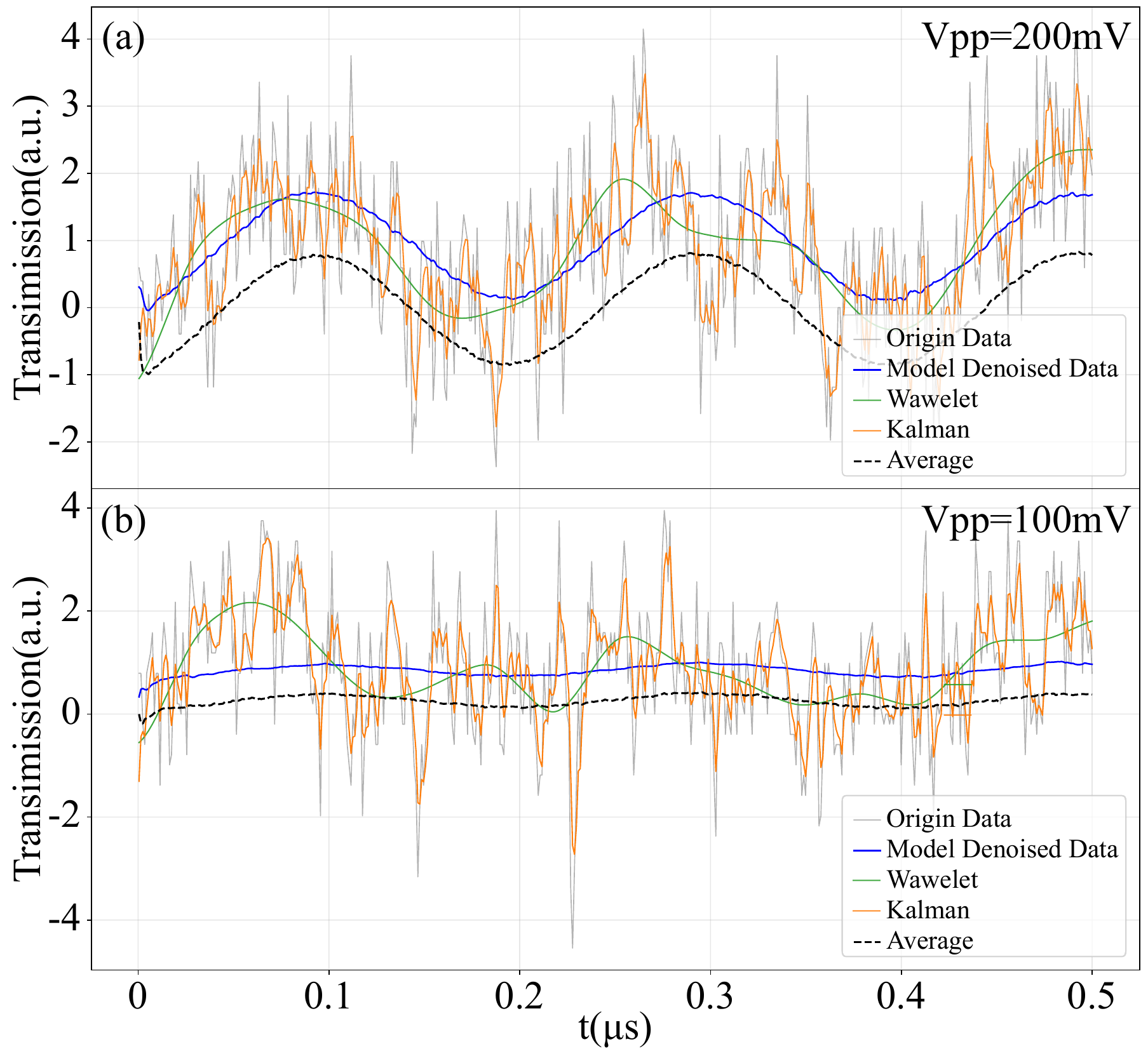}
    \caption{Different denoising models on time-domain signals for Vpp=200 mV (a) and 100 mV (b). Adopted denoising methods include wavelet denoising (green solid line), Kalman filtering (orange solid line), multi-measurement averaging (black dashed line), and the proposed model’s denoising results (blue solid line), all applied to the single-shot measurement result of the probe beam transmission signal (gray solid line). Parameters of wavelet denoising and Kalman filtering were optimized by minimizing the error between their denoising results and the multi-measurement averaging result, as detailed in the main text.}
    \label{fig:time}
\end{figure}

\subsection{Quantitative Validation of Denoising Performance: Deep Learning vs. Traditional Methods}

Fig. \ref{fig:time} validates the proposed deep learning model’s superiority over traditional methods (average against 10000 sets, wavelet transform and Kalman filter) in suppressing noise in time series (recorded by the oscilloscope), using IF time-domain signals from Rydberg atom microwave heterodyne detection.

\begin{table*}
    \centering
\caption{MSE for different denoising methods on time domain data for Vpp= 200, 100 mV cases.}
\label{tab:timeMSE}
    \begin{tabular}{|c|c|c|c|}\hline
         &  Deep learning&  Kalman filtering& Wavelet\\\hline
         Vpp=200 mV&  \(\mathbf{7.71\times10^{-4}\pm1.5\times10^{-4}}\)&  \(7.0\times10^{-2}\pm1.6\times10^{-2}\)&  \(7.1\times10^{-2}\pm2.5\times10^{-2}\)\\\hline
         Vpp=100 mV&  \(\mathbf{7.9\times10^{-5}\pm4.5\times10^{-5}}\)&  \(1.4\times10^{-1}\pm3.6\times10^{-2}\)& \(2.5\times10^{-1}\pm6.4\times10^{-2}\)\\ \hline
    \end{tabular}

\end{table*}

Wavelet transform denoiseing\cite{HUIMIN20121354,FANG200467} (green solid line) used coiflets wavelet (4th type) with a threshold of 0.55. 
Kalman filtering\cite{5311910} (orange solid line) parameters were set as: process noise variance \(Q = 0.0175\) and measurement noise variance \(R = 0.06\). 
To make a fair comparison, the above parameters of traditional methods were optimized by grid search to minimize the MSE between denoised data by these methods and the averaged result of 10,000 repeated measurements with the same IF signal and i.i.d. noise.

The 10,000-measurement average (black dashed line) was visually smoother than the single-shot result (gray solid line), with noise suppressed and IF signal prominent. 
The deep learning model’s denoised results (blue solid line) were closer to this average than traditional methods, yielding smoother curves for noisy single-frequency IF signals with three orders of magnitude high temporal efficiency.

For different voltage on the dipole plates, 200 mV,  Fig.~\ref{fig:time} (a), and 100 mV,  Fig.~\ref{fig:time} (b), the MSE to averaging result  (mean$\pm$ standard deviation from 2000 distinct time-domain signals in the test set)   for these three methods is shown in Table \ref{tab:timeMSE}.
This two-order-of-magnitude difference in MSE confirms the deep learning model's significant denoising advantage.

\begin{figure*}
    \centering
    \includegraphics[width=1\linewidth]{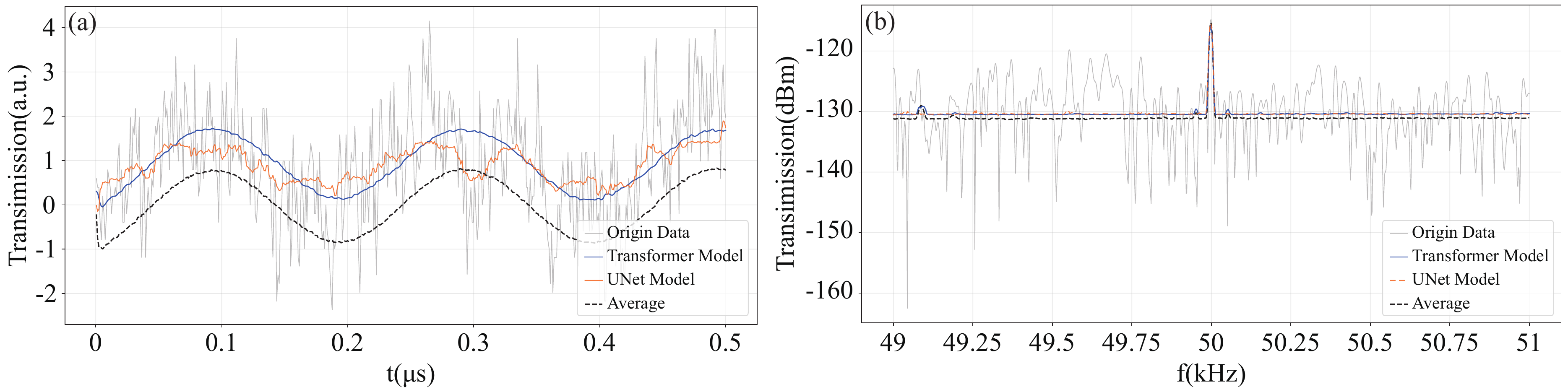}
    \caption{Denoising results of different deep learning models under the same self-supervised architecture. (a) Denoising results of Transformer (blue solid line) and convolution based U-Net structure (orange solid line), both applied to the single-shot measurement result of the probe beam transmission signal (gray solid line); the averaged result of 10,000 measurements (black dashed line) serves as the Ground truth reference. (b) Denoising results of Transformer (blue solid line) and convolution based U-Net structure (orange dashed line) applied to the frequency-domain data of the single-shot measurement result (gray solid line). The Transformer result exhibits not only the intermediate frequency (IF) signal but also its sidebands and the weak ~49.1 kHz signal, and is closer to the averaged result than the U-Net.}
    \label{fig:compare}
\end{figure*}

\subsection{Relationship Between Model Complexity and Denoising Performance}

To investigate the influence of model complexity on denoising performance, Fig. \ref{fig:compare} compares two deep learning architectures (i.e., Transformer- vs. U-net model\cite{ronneberger2015u,9059879,IBTEHAZ202074}, structure disgram shown in Appendix \ref{sec:model}) on noisy time- and frequency-domain data.
The Transformer model (U-net model) has parameter sizes of $4.50\times 10^5$ ($1.59\times 10^4$) and training times of 2000 s per epoch (10 s per epoch), respectively; other parameters (e.g., epochs, dataset configuration, and optimizer ) were kept consistent.

In both domains, the denoised results of the Transformer-based model are closer to the 10,000-measurement average than those of its counterpart.
For the time domain: the alternative architecture fails to capture IF signal features, resulting in outputs  that contain significant high-frequency noise alongside the 50 kHz oscillation. In contrast, the Transformer model identifies and extracts the shared IF signal while suppressing noise interference. 
For the frequency domain: only the IF signal is prominent in the results of the alternative architecture. 
However, the Transformer model retains the IF signal, its sidebands, and the weak 49.1 kHz signal, showing higher similarity to the 10,000-measurement average.
These results demonstrate that increased computational complexity (quantified by training parameters) enhances weak signal extraction, despite longer training times. 
This enables a more robust recovery of weak signals from noise; future work may adopt deeper or higher-complexity architectures to further enhance performance.

\section{Discussion}

In this work, we propose a self-supervised deep learning framework for denoising in Rydberg atom microwave sensors. 
This framework achieves denoising performance comparable to the averaging method both in the frequency and time domains of microwave sensing signals, while exhibiting extremely low latency. 
It reduces processing time by three orders of magnitude compared to the averaging method. 
Compared with Kalman filtering and wavelet transform-based denoising methods, the proposed approach demonstrates a superior denoising effect. 
Furthermore, by replacing the computational units within the deep learning framework, we identify an optimization direction for further performance enhancement, namely increasing the model's computational complexity.

The performance of the deep learning denoising framework stems inherently from the model's ability to learn the intrinsic properties of the target signals \cite{izadi2023image}. 
Rydberg atom microwave signals are constrained by the physical laws governing energy level transitions, exhibiting well-defined underlying structures in their amplitude variations, frequency distributions, and temporal evolution. 
In contrast, environmental interference and quantum fluctuations manifest as random noise components that follow specific statistical laws. 
By extracting features of common signals from a large dataset, the model can leverage this prior knowledge to retrieve stable shared components from single noisy measurements while suppressing random noise variations across different samples. 
This process is equivalent to achieving ``virtual averaging'' in a data-driven manner, preserving the temporal resolution of single-shot measurements without sacrificing the speed limitations associated with traditional averaging methods.

Compared to conventional techniques, the proposed deep learning-based denoising method for quantum sensing offers three distinct advantages.
First, the model inference process is extremely fast, with single data processing completed within milliseconds, fully meeting the speed requirements for dynamic process monitoring. 
Second, centered on data-driven learning, the method eliminates the need for established noise models, enabling stronger adaptability to non-stationary noise commonly encountered in open environments. 
Third, model training can be accomplished without the need for clean reference signals, making it suited for scenarios in the quantum sensing field where labeled data are scarce.

Future research will focus on the following limitations: for the data-driven paradigm, model performance is highly dependent on the quality of the training data set, and sufficient data covering various noise scenarios are prerequisite for ensuring denoising effectiveness. 
Additionally, although the Transformer architecture delivers superior denoising performance, its computational complexity poses challenges for deployment in embedded sensing systems with limited computing resources.

\section{Acknowledgment}

We acknowledge the the National Key R\&D Program of China (grant no. 2022YFA1404003), the National Natural Science Foundation of China (grants T2495252, 12104279, 123B2062, 12574318), Innovation Program for Quantum Science and Technology (Grant No. 2021ZD0302100),  the Fund for Shanxi ‘1331 Project’ Key Subjects Construction.

\section{Data availability}

The datasets generated and/or analyzed during the current study are available in the Zenodo repository, accessible via the following Digital Object Identifier (DOI):10.5281/zenodo.18091258

\newpage
\appendix

\section{Model structure}\label{sec:model}

\begin{figure*}
    \centering
    \includegraphics[width=1\linewidth]{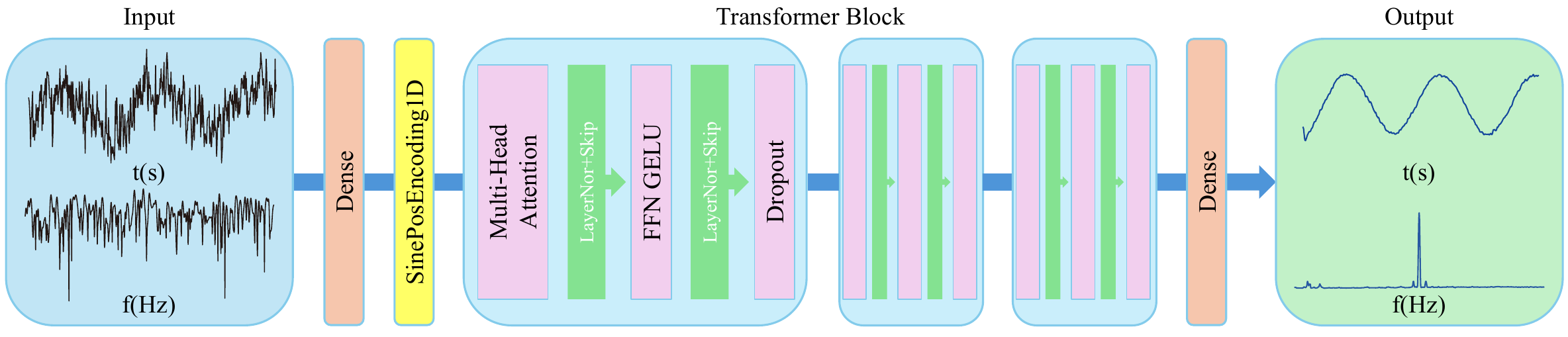}
    \caption{Transformer based mode structure.}
    \label{fig:transformer}
\end{figure*}

\begin{figure}
    \centering
    \includegraphics[width=1\linewidth]{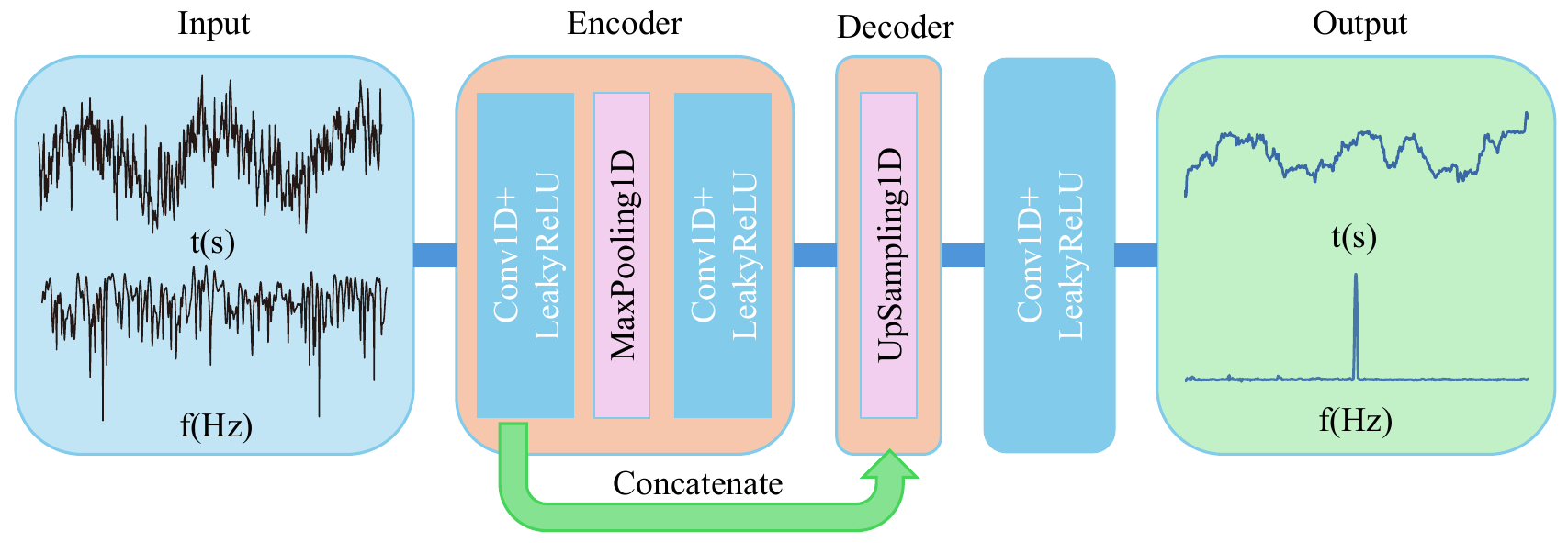}
    \caption{U-Net structure.}
    \label{fig:Unet}
\end{figure}

The structure of transformer model and U-Net model are shown in Figs. \ref{fig:transformer} and \ref{fig:Unet}, respectively.

The structure of transformer model is follows. First, a Dense layer projects the demension of data to a 64-dimension space for feature mapping.
Second, sine encoding layer (SinePosEncoding) assigns unique vectors via sine values to preserve sequence order.
Then, three stacked encoder blocks (each with 8-head multi-head self attention for local-global dependencies, 128-dimension feed-ford neural network with a Gaussian Error Linear Unit (GELU) activation function for signal representation, residual connections for fusion, and dropout (with rate \(p=0.1\))).
Finally, a Dense layer maps 64-dimension features back to spectral intensity for denoising.
Weights of the above layers were updated via backpropagation by the Adam optimizer (with parameter decay coefficient for  first-order momentum  \(\beta_1=0.9\), decay coefficient for  second-order momentum \(\beta_2=0.999\),  stability parameter that prevents division by zero\(\epsilon=10^{-7}\), and learning rate $1.0\times10^{-5}$) with mean square error (MSE) loss (\(\mathcal{L}\)) between predictions and \(y_{\text{train}}\). The hardware for the deep learning model is a CPU with the parameters: 12th Gen Intel(R) Core(TM) i7-12700 (2.10 GHz).

The U-Net model comprises an encoder, a decoder, and a 1D convolutional layer, where the encoder and decoder are connected via layer concatenation.
The encoder consists of three components. The first convolutional layer contains 24 convolution kernels, followed by a LeakyReLU activation function. Subsequently, a max-pooling layer with a pooling size of 2 is applied to halve the temporal resolution. The second convolutional layer further extracts high-level temporal features at the downsampled temporal scale.
The decoder restores temporal resolution using a 1D upsampling layer with a scale factor of 2. Via skip connections, the upsampled features are concatenated with the corresponding high-resolution features from the encoder, forming the characteristic U-shaped network structure. This design enables the network to effectively fuse low-level temporal detail information and high-level semantic features.
Finally, a 1D convolutional layer (with 48 convolution kernels and a kernel size of 3) is used to fuse multi-scale features, accompanied by the LeakyReLU activation function.

\end{document}